\documentclass[11pt]{spie} 

\newcommand{\fourIdx}[5]{%
 \setbox1=\hbox{\ensuremath{^{#1}}}%
 \setbox2=\hbox{\ensuremath{_{#2}}}%
 \setbox5=\hbox{\ensuremath{#5}}%
 \hspace{\ifnum\wd1>\wd2\wd1\else\wd2\fi}%
 \ensuremath{\copy5^{\hspace{-\wd1}\hspace{-\wd5}#1\hspace{\wd5}#3}%
 _{\hspace{-\wd2}\hspace{-\wd5}#2\hspace{\wd5}#4}%
 }}

\usepackage[utf8]{inputenc} 

\usepackage{geometry} 
\usepackage{graphicx} 

\usepackage{booktabs} 
\usepackage{array} 
\usepackage{amsmath} 
\usepackage{paralist} 
\usepackage{verbatim} 
\usepackage{subfig} 

\usepackage{fancyhdr} 
\usepackage{sectsty}
\allsectionsfont{\sffamily\mdseries\upshape} 

\usepackage[nottoc,notlof,notlot]{tocbibind} 
\usepackage[titles,subfigure]{tocloft} 

\newcommand{\ud}{\mathrm{d}}
\newcommand{\uD}{\mathrm{D}}
\newcommand{\uuD}{\mathcal{D}}

\title{Path Integral by Space-time Slicing Approximation In Open Bosonic String Field} 
\author{Am-Gil Ri\supit{a}, Tae-Song Kim\supit{a}, Chol-Man Ri\supit{b} and Song-Jin Im\supit{c}\skiplinehalf
	\supit{a} Faculty of Energy Science, Kim Il Sung University, Pyongyang, DPR Korea\skiplinehalf
	\supit{b} E-Library, Kim Il Sung University, Pyongyang, DPR Korea\skiplinehalf
	\supit{c} Faculty of Physics, Kim Il Sung University, Pyongyang, DPR Korea
}

\authorinfo{Kim Tae-Song:E-mail: ryongnam13@yahoo.com}

\date{} 

\begin{document} 
\maketitle 
\begin{abstract}
 In our paper, we considered how to apply the traditional Feynman path integral tostring field. By constructing the complete set in Fock space of non-relativistic and relativistic open bosonic string fields, weextended Feynman path integral to path integral on functional field and use it to quantize open bosonic string field.
\end{abstract} 


\section{Introduction}
  String field theories come in a number of varieties depending on which type of string is second quantized: open string field theories describe the scattering of open strings, closed string field theories describe closed strings, while open-closed stringfield theories include both open and closed strings. 

In addition, depending on the method used to fix the worldsheet diffeomorphisms and conformal transformations in the original free string theory, the resulting string field theories can be very different. Using light cone gauge yields light-cone string field theories whereas using BRST quantization, one finds covariant string field theories.\cite{ref1}

There are also hybrid string field theories, known as covariantized light-cone string field theories which use elements of both light-cone and BRST gauge-fixed string field theories.\cite{ref1} 

A final form of string field theory, known as background independent open string field theory, takes a very different form; instead of second quantizing the worldsheet string theory, it second quantizes thespace of two-dimensional quantum field theories.\cite{ref2}

In our paper we will deal with the light-cone string field theory by extending the traditional path integral.

Light-cone string fieldtheories were introduced by Stanley Mandelstam\cite{ref3} and developedby Mandelstam, Michael Green, John Schwarz and Lars Brink.\cite{ref4} 

Anexplicit description of the second-quantization of the light-cone string was given by Michio Kaku and Keiji Kikkawa.\cite{ref5}

In 1948, R. P. Feynman\cite{ref6} expressed the integral kernel of the fundamental solution for the Schrödinger equation, using the path integral. 

Feynman explained his path integral as a limit ofa finite dimensional integral, which is now called the time slicing approximation. Furthermore, Feynman considered path integrals with general functional $F[\gamma]$ as integrand, and suggested a new analysis on a path space with the functional integration $\int F[\gamma] e^{\frac{i}{\hbar} S[\gamma]}\uuD[\gamma]$ and the functional differentiation $(DF)[\gamma][\eta]$.

However, in 1960, R. H. Cameron\cite{ref7} provedthat the measure $e^{\frac{i}{\hbar}S[\gamma]} \uuD[\gamma]$ of Feynman path integrals does not exist in mathematics. 

After that, using the time slicing approximation, Naoto and Fujiwara\cite{ref8} proved the existence of the Feynman path integrals $\int F[\gamma] e^{\frac{i}{\hbar} S[\gamma]}\uuD[\gamma]$ with the smooth functional derivatives $(DF)[\gamma][\eta]$. 

In our paper we will represent the open bosonic string field by generalizing Feynman path integral from functional Schrödinger equation of it, mostly repeating the same progresses as Feynman expressed the fundamental solution of Schrödinger equation of point particle with his path integral.

\section{Path Integral Quantization of String Field}

The classical string field functionals satisfy the string Schrodinger equation as follows.\cite{ref9} 
\begin{equation}
\label{eq:1} \frac{\partial}{\partial
t}\Phi[\mathbf{X}]=\frac{1}{2T}\int \ud\sigma\left(-\frac{\delta^2}{(\delta\mathbf{X}(\sigma)^2}+T^2\mathbf{X}^\prime (\sigma)^2\right)=\hat{h}\Phi[\mathbf{X}]
\end{equation}
The free Lagrangian resulting in this equation ofopen bosonic string field is: \begin{equation} \label{eq:2} L_0=\frac{1}{2}\int \uD \mathbf{X}\Phi^+[\mathbf{X}]\left(i \frac{\partial}{\partial t}-\hat{h}\right)\Phi[\mathbf{X}]+h.c.
\end{equation} 
, where $h.c.$ means Hermitian conjugation. By means of path integral by space-time slicing approximation, we’ll obtain the formal solution of Schrodinger equation of quantum string field 
\begin{equation} 
\label{eq:3}
i\frac{\partial}{\partial t}\left|\Psi(t)\right>=\hat{H}^{(2)}\left |\Psi(t)\right>
\end{equation}
, where $\hat{H}^{(2)}$ is follows.
\begin{equation} 
\label{eq:4} 
\hat{H}^{(2)}=\frac{1}{2}\int \uD\mathbf{X}(\sigma)\hat{\Phi}^+[\mathbf{X}]\hat{h}\hat{\Phi}[\mathbf{X}] 
\end{equation} 
The eigenvalue problem of string field functional operator is set as follows.
\begin{equation} 
\label{eq:5} 
\hat{\Phi}[\mathbf{X}]\left|\Phi\right>=\Phi[\mathbf{X}]\left|\Phi\right> 
\end{equation}
Inorder to get the complete condition, we use the complete set in Fock space as 
\begin{equation} 
\label{eq:6}
\sum_{n=0}^\infty{\int \uD\mathbf{X}_1\cdots\uD\mathbf{X}_n\left|\mathbf{X}_1,\cdots,\mathbf{X}_n\right>\left<\mathbf{X}_n,\cdots,\mathbf{X}_1\right|}=I 
\end{equation}
,where 
\begin{equation}
\label{eq:7}
\left|\mathbf{X}_1,\cdots,\mathbf{X}_n\right>=\frac{1}{\sqrt{n!}}\hat{\Phi}^+[\mathbf{X}_1]\cdots\hat{\Phi}^+[\mathbf{X}_n ]\left|0\right>
\end{equation}
Using the complete set (\ref{eq:6}), we get 
\begin{equation}
\label{eq:8}
\left|\Phi\right>=\sum_{n=0}^\infty{\int\uD\mathbf{X}_1\cdots\uD\mathbf{X}_n\left|\mathbf{X}_1,\cdots,\mathbf{X}_n\right>\left<\mathbf{X}_n,\cdots,\mathbf{X}_1\middle|\Phi\right>}
\end{equation}
To determine $\left<\mathbf{X}_n,\cdots,\mathbf{X}_1\middle|\Phi\right>$, if we repeat to use 
\begin{equation} 
\label{eq:9}
\left<\mathbf{X}_n,\cdots,\mathbf{X}_1\middle|\Phi\right>=\frac{1}{\sqrt{n}}\left<\mathbf{X}_{n-1},\cdots,\mathbf{X}_1\middle|\hat{\Phi}[\mathbf{X}_n]\middle|\Phi\right>=\frac{1}{\sqrt{n}}\left<\mathbf{X}_{n-1},\cdots,\mathbf{X}_1\middle|\Phi\right>\Phi[\mathbf{X}_n] 
\end{equation}
we find 
\begin{equation*}
\left<\mathbf{X}_n,\cdots,\mathbf{X}_1\middle|\Phi\right>=\frac{1}{\sqrt{n!}}\left<0\middle|\Phi\right>\Phi[\mathbf{X}_n]\cdots\Phi[\mathbf{X}_1]
\end{equation*} 
In such way, we can see 
\begin{equation*}
\left<\Phi\middle|\mathbf{X}_n,\cdots,\mathbf{X}_1\right>=\frac{1}{\sqrt{n!}}\Phi^+[\mathbf{X}_n]\cdots\Phi^+[\mathbf{X}_1]\left<\Phi\middle|0\right>
\end{equation*}
Now if we assume $\left<\Phi\middle|0\right>=\left<0\middle|\Phi\right>=1$, we get 
\begin{equation} 
\label{eq:10}
\left<\mathbf{X}_n,\cdots,\mathbf{X}_1\middle|\Phi\right>=\frac{1}{\sqrt{n!}}\Phi[\mathbf{X}_n]\cdots\Phi[\mathbf{X}_1], 
\end{equation} 
\begin{equation} 
\label{eq:11}
\left<\Phi\middle|\mathbf{X}_n,\cdots,\mathbf{X}_1\right>=\frac{1}{\sqrt{n!}}\Phi^+[\mathbf{X}_n]\cdots\Phi^+[\mathbf{X}_1].
\end{equation}
Then, substituting the equation (\ref{eq:8}) for (\ref{eq:10}), we obtain 
\begin{equation} 
\label{eq:12} 
\begin{split}
\left|\Phi\right>&=\sum_{n=0}^\infty{\int \uD \mathbf{X}_1\cdots\uD\mathbf{X}_n\left|\mathbf{X}_1,\cdots,\mathbf{X}_n\right> \left<\mathbf{X}_n,\cdots,\mathbf{X}_1\middle|\Phi\right>}\\
&=\sum_{n=0}^\infty{\int\uD\mathbf{X}_1\cdots\uD\mathbf{X}_n\frac{1}{\sqrt{n!}}\Phi[\mathbf{X}_n]\cdots\Phi[\mathbf{X}_1]\left|\mathbf{X}_1,\cdots,\mathbf{X}_n\right>}\\
&=\sum_{n=0}^\infty{\int\uD\mathbf{X}_1\cdots\uD\mathbf{X}_n\frac{1}{\sqrt{n!}}\Phi[\mathbf{X}_n]\cdots\Phi[\mathbf{X}_1]\hat{\Phi}^+[\mathbf{X}_1]\cdots\hat{\Phi}^+[\mathbf{X}_n]\left|0\right>}\\
&=e^{\int\uD\mathbf{X}\Phi[\mathbf{X}]\hat{\Phi}^+[\mathbf{X}]}\left|0\right>
\end{split} 
\end{equation} 
Similarly, we can easily find
\begin{equation} 
\label{eq:13}
\left<\Phi\right|=\left<0\right|e^{\int\uD\mathbf{X}\hat{\Phi}[\mathbf{X}]\Phi^+ [\mathbf{X}]}. 
\end{equation} 
Using this, we can express the scalar product between two state vectors as 
\begin{equation}
\label{eq:14} 
\begin{split}
\left<\Phi'\middle|\Phi''\right>&=\left<0\middle|e^{\int\uD\mathbf{X}\hat{\Phi}'[\mathbf{X}] \Phi'^+[\mathbf{X}]}e^{\int\uD\mathbf{X}\Phi''[\mathbf{X}]\hat{\Phi}''^+[\mathbf{X}]}\middle|0\right>\\
&=\sum_{n=0}^\infty{\int\uD\mathbf{X}_1\cdots\mathbf{X}_n\frac{1}{n!}\left<\Phi'\middle|\mathbf{X}_1,\cdots,\mathbf{X}_n\right>\left<\mathbf{X}_n,\cdots,\mathbf{X}_1\middle|\Phi''\right>}\\
&=\sum_{n=0}^\infty{\int\uD\mathbf{X}_1\cdots\mathbf{X}_n\frac{1}{n!}\Phi'^+[\mathbf{X}_n]\cdots\Phi'^+[\mathbf{X}_1]\Phi''[\mathbf{X}_n]\cdots\Phi''[\mathbf{X}_1]}\\
&=e^{\int\uD\mathbf{X}\Phi'^+[\mathbf{X}] \Phi''[\mathbf{X}]}
\end{split}
\end{equation}
 Now we can write the complete set of space as 
\begin{equation} 
\label{eq:15}
\int\tilde{\uD}\Phi^+\tilde{\uD}\Phi\left|\Phi\right>\left<\Phi\right|=I
\end{equation}
, where 
\begin{equation} 
\label{eq:16}
\tilde{\uD}\Phi^+\tilde{\uD}\Phi=\prod_{\mathbf{X}(\sigma)}{\uD\Phi^+[\mathbf{X}]\uD\Phi[\mathbf{X}]e^{-\int\uD\mathbf{X}\Phi^+[\mathbf{X}]\Phi[\mathbf{X}]}}
\end{equation}
Here
\begin{equation*}
\uD\Phi[\mathbf{X}]\equiv\lim_{N\rightarrow\infty}\prod_{\{\mathbf{X}(\sigma_1),\cdots,\mathbf{X}(\sigma_N)\}}\ud\Phi(\mathbf{X}(\sigma_1),\cdots,\mathbf{X}(\sigma_N)).
\end{equation*}
It is that string field functional can be considered as a function of infinite number of variables. Takingaccount for the equations (\ref{eq:12}), (\ref{eq:13}) and (\ref{eq:16}), the equation (\ref{eq:15}) rewrite as
\begin{equation*}
\begin{split}
\int\tilde{\uD}\Phi^+\tilde{\uD}\Phi\left|\Phi\right>\left<\Phi\right|&=\int\prod_{\mathbf{X}}\uD\Phi^+[\mathbf{X}]\uD\Phi[\mathbf{X}] e^{-\int\uD\mathbf{X}\Phi^+[\mathbf{X}]\Phi[\mathbf{X}]} e^{\int\uD\mathbf{X}\Phi[\mathbf{X}]\hat{\Phi}^+[\mathbf{X}]}\left|0\right> \left<0\right|e^{\int\uD\mathbf{X}\hat{\Phi}[\mathbf{X}]\Phi^+[\mathbf{X}]}\\
&=\sum_{k,l=0}^\infty\frac{1}{\sqrt{k!l!}}\int\prod_{\mathbf{X}}\uD\Phi^+[\mathbf{X}]\uD\Phi[\mathbf{X}] e^{-\int\uD\mathbf{X}\Phi^+[\mathbf{X}]\Phi[\mathbf{X}]} \int\uD\mathbf{X}_1\cdots\uD\mathbf{X}_k\\ 
&\qquad\qquad\uD\mathbf{Y}_1\cdots\uD\mathbf{Y}_l \Phi[\mathbf{X}_1]\cdots\Phi[\mathbf{X}_k]\Phi^+[\mathbf{Y}_1]\cdots\Phi^+[\mathbf{Y}_l] \left|\mathbf{X}_1,\cdots,\mathbf{X}_k\right>\left<\mathbf{Y}_1,\cdots,\mathbf{Y}\right|
\end{split}
\end{equation*}

Now taking account for
 \begin{equation*}
\begin{split}
&\int\uD\Phi^+\uD\Phi e^{-\int \uD \mathbf{X}\Phi^+[\mathbf{X}]\Phi[\mathbf{X}]}\\
&\qquad \qquad\int \uD\mathbf{X}_1\cdots\uD\mathbf{X}_k\uD\mathbf{Y}_1\cdots\uD\mathbf{Y}_l \Phi[\mathbf{X}_1]\cdots\Phi[\mathbf{X}_k]\Phi^+[\mathbf{Y}_1]\cdots\Phi^+[\mathbf{Y}_l] F\left[\mathbf{X}_1,\cdots,\mathbf{X}_k \middle| \mathbf{Y}_1,\cdots,\mathbf{Y}_l\right] \\
&=\delta_{kl}k!F\left[\mathbf{X}_1,\cdots,\mathbf{X}_k \middle| \mathbf{Y}_1,\cdots,\mathbf{Y}_k\right]
\end{split}
\end{equation*}
, we can find that the above equation is 
 \begin{equation*}
\int\tilde{\uD}\Phi^+\tilde{\uD}\Phi \left|\Phi\right>\left<\Phi\right|=\sum_{k=0}^\infty \int\uD\mathbf{X}_1\cdots\mathbf{X}_k\left|\mathbf{X}_1,\cdots,\mathbf{X}_k\right>\left<\mathbf{X}_k,\cdots,\mathbf{X}_1\right| = I.
\end{equation*}
Next, we introduce $\delta-$functional. Substituting the complete set (\ref{eq:15}) for the scalar product, we get
\begin{equation*}
\begin{split}
\left<\Phi'\middle| \Phi''\right>&=\int \tilde{\uD}\Phi^+\tilde{\uD}\Phi\left<\Phi'\middle|\Phi\middle> \middle<\Phi \middle| \Phi''\right>\\
&=\int \uD\Phi^+\uD\Phi e^{-\int \uD\mathbf{X}\Phi^+[\mathbf{X}]\Phi[\mathbf{X}]} e^{\int \uD\mathbf{X}\Phi'^+[\mathbf{X}]\Phi[\mathbf{X}]} e^{\int \uD\mathbf{X}\Phi^+[\mathbf{X}]\Phi''[\mathbf{X}]}\\
&=\int \uD\Phi^+e^{-\int \uD\mathbf{X}\Phi^+[\mathbf{X}](\Phi[\mathbf{X}]-\Phi''[\mathbf{X}])} \int \uD\Phi e^{\int \uD\mathbf{X}\Phi'^+[\mathbf{X}]\Phi[\mathbf{X}]}
\end{split}
\end{equation*}
By the way in order that this is agreed with the equation (\ref{eq:14}), the following relation must be satisfied.
\begin{equation}
\label{eq:17}
\delta[\Phi'-\Phi'']=\int \uD\Phi^+ e^{-\int \uD\mathbf{X}\Phi^+[\mathbf{X}](\Phi'[\mathbf{X}]-\Phi''[\mathbf{X}])} =\int\prod_{\mathbf{X}}\uD\Phi^+[\mathbf{X}] e^{-\int \uD\mathbf{X}\Phi^+[\mathbf{X}](\Phi'[\mathbf{X}]-\Phi''[\mathbf{X}])} 
\end{equation}
Now introducing the representation as
\begin{equation}
\label{eq:18}
\begin{split}
\fourIdx{*}{}{}{}{\left<\Phi\right|}=e^{-\frac{1}{2}\int\uD \mathbf{X}\Phi^+[\mathbf{X}]\Phi[\mathbf{X}]} \left<\Phi\right|\\
\left|\Phi\right>^*=\left|\Phi\right> e^{-\frac{1}{2}\int\uD \mathbf{X}\Phi^+[\mathbf{X}]\Phi[\mathbf{X}]}
\end{split}
\end{equation}
, the complete set (\ref{eq:15}) can be rewritten as
\begin{equation}
\label{eq:19}
\int\uD \Phi^+\uD \Phi \left|\Phi\right>^*\fourIdx{*}{}{}{}{\left<\Phi\right|} = I
\end{equation}
In this representation the scalar product is expressed as 
\begin{equation}
\label{eq:20}
\begin{split}
\fourIdx{*}{}{*}{}{\left<\Phi'\middle|\Phi''\right>}&= e^{-\frac{1}{2}\int\uD \mathbf{X}\Phi'^+[\mathbf{X}]\Phi'[\mathbf{X}]}\left<\Phi'\middle|\Phi''\right> e^{-\frac{1}{2}\int\uD \mathbf{X}\Phi''^+[\mathbf{X}]\Phi''[\mathbf{X}]}\\
&=e^{-\frac{1}{2}\int\uD \mathbf{X}\{\Phi'^+[\mathbf{X}](\Phi'[\mathbf{X}]-\Phi''[\mathbf{X}]) + (\Phi''^+[\mathbf{X}]-\Phi'^+[\mathbf{X}])\Phi''[\mathbf{X}]\}}
\end{split}
\end{equation}
Then we can obtain the traditional path integral representation of the transition amplitude of string field in Schrödinger picture.
To do so, introducing the evolution operator that satisfies the following
\begin{equation*}
\left|\Psi(t)\right>=U(t,t_0)\left|\Psi(t_0)\right>
\end{equation*}
, this operator is formally got as follows.
\begin{equation*}
U(t,t_0)=TP\{e^{-i\hat{H}^{(2)}(t-t_0)}\}
\end{equation*} 
Here $T$ means time-ordered product and $P$ means space-ordered product.
Then the solution can be represented as
\begin{equation}
\label{eq:21}
\left|\Psi(t)\right>=TP\{ e^{-i\hat{H}^{(2)}(t-t_0)} \}\left|\Psi(t_0)\right>
\end{equation}
We factorize both sides of this equation into the scalar product $\fourIdx{*}{}{}{}{\left<\Phi\right|}$ to both sides of this equation. Then 
\begin{equation*}
\begin{split}
\fourIdx{*}{}{}{}{\left<\Phi\middle|\Psi(t)\right>}&=\fourIdx{*}{}{}{}{\left<\Phi\middle|TP\{ e^{-i\hat{H}^{(2)}(t-t_0)} \}\middle|\Psi(t_0)\right>}\\
&=\int\uD\Phi^+_0\uD\Phi_0 \fourIdx{*}{}{}{}{\left<\Phi\right|}TP\{ e^{-i\hat{H}^{(2)}(t-t_0)} \}\fourIdx{}{}{*}{}{\left|\Phi_0\right>} \fourIdx{*}{}{}{}{\left<\Phi_0\middle|\Psi(t_0)\right>}
\end{split}
\end{equation*} 
Let’s divide the time interval $(t-t_0)$ into infinite segments each lasting
\begin{equation*}
\Delta t=\frac{t-t_0}{n},   (n\rightarrow\infty)
\end{equation*}
and insert the complete sets between all these factors $TP\{e^{-i\Delta t \hat{H}^{(2)}}\}$. Then the kernel is written 
\begin{equation}
\label{eq:22}
\fourIdx{*}{}{*}{}{\left<\Phi\middle|TP\{ e^{-i\hat{H}^{(2)}(t-t_0)} \}\middle|\Phi_0\right>}=\lim_{n\rightarrow\infty}\int\prod_{j=1}^{n-1}\uD\Phi^+_j\uD\Phi_j \prod_{j=0}^{n-1} \fourIdx{*}{}{*}{}{\left<\Phi_{j+1}\middle|TP\{ e^{-i\Delta t\hat{H}^{(2)}} \}\middle|\Phi_j\right>}
\end{equation}
Since $\Delta t$ is fairly small, the following is satisfied.
\begin{equation*} 
\begin{split}
\fourIdx{*}{}{}{}{\left<\Phi_{j+1}\middle|TP\{ e^{-i\Delta t\hat{H}^{(2)}} \}\middle|\Phi_j\right>}&=\fourIdx{*}{}{*}{}{\left<\Phi_{j+1}\middle|1-i\Delta t\hat{H}+O(\Delta t^2)\middle|\Phi_j\right>}\\
&=\fourIdx{*}{}{*}{}{\left<\Phi_{j+1}\middle|\Phi_j\right>}-i\Delta t~ \fourIdx{*}{}{*}{}{\left<\Phi_{j+1}\middle|\int\uD\mathbf{X}\hat{\Phi}^+[\mathbf{X}]\hat{h}\hat{\Phi}[\mathbf{X}]\middle|\Phi_j\right>}+\cdots\\
&=\fourIdx{*}{}{*}{}{\left<\Phi_{j+1}\middle|\Phi_j\right>}-i\Delta t~ \fourIdx{*}{}{*}{}{\left<\Phi_{j+1}\middle|\Phi_j\right>} \int\uD\mathbf{X}\Phi^+_{j+1}[\mathbf{X}]\hat{h}\Phi_j[\mathbf{X}]+\cdots\\
&=\fourIdx{*}{}{*}{}{\left<\Phi_{j+1}\middle|\Phi_j\right>} e^{-i\Delta t H^{cl}(\Phi_{j+1},\Phi_j)}\\
&=e^{-i\Delta t H^{cl}(\Phi_{j+1},\Phi_j)} e^{-\frac{1}{2}\int\uD\mathbf{X}\left\{\Phi_{j+1}^+[\mathbf{X}](\Phi_{j+1}[\mathbf{X}]-\Phi_j[\mathbf{X}])+(\Phi^+_j[\mathbf{X}]-\Phi^+_{j+1}[\mathbf{X}])\Phi_j[\mathbf{X}]\right\}}
\end{split}
\end{equation*}
, where $H^{(cl)}$ is the Hamiltonian of classical string field. Substituting this for the equation (\ref{eq:22}), 
 \begin{equation*}
\begin{split}
&\fourIdx{*}{}{*}{}{\left<\Phi\middle|TP\{ e^{-i\hat{H}^{(2)}(t-t_0)} \}\middle|\Phi_0\right>}=\lim_{n\rightarrow\infty}\int\nolimits\prod_{j=1}^{n-1}\uD\Phi^+_j \uD\Phi_j \prod_{j=0}^{n-1}\fourIdx{*}{}{*}{}{\left<\Phi_{j+1}\middle|TP\{ e^{-i\Delta t\hat{H}^{(2)}} \}\middle|\Phi_j\right>}\\
&=\lim_{n\rightarrow\infty}\int\prod_{j=1}^{n-1}\uD\Phi_j^+ \uD\Phi_j \prod_{j=0}^{n-1} e^{-i\Delta t H^{(cl)}(\Phi_{j+1},\Phi_j)} e^{-\frac{1}{2}\int\uD\mathbf{X}\{\Phi_{j+1}^+[\mathbf{X}](\Phi_{j+1}[\mathbf{X}]-\Phi_j[\mathbf{X}])+(\Phi_j^+[\mathbf{X}]-\Phi_{j+1}^+[\mathbf{X}])\Phi_j[\mathbf{X}]\} }\\
&=\lim_{n\rightarrow\infty}\int\prod_{j=1}^{n-1}\uD\Phi_j^+ \uD\Phi_j e^{i\Delta t \sum_{j=0}^{n-1}\int\uD\mathbf{X}\left(\frac{i}{2}\Phi_{j+1}^+[\mathbf{X}]\frac{\Phi_{j+1}[\mathbf{X}]-\Phi_j[\mathbf{X}]}{\Delta t}-\frac{i}{2}\frac{\Phi_{j+1}^+[\mathbf{X}]-\Phi_j^+[\mathbf{X}]}{\Delta t}\Phi_j[\mathbf{X}]-\Phi_{j+1}^+[\mathbf{X}]\hat{h}\Phi_j[\mathbf{X}]\right)}\\
&=\lim_{n\rightarrow\infty}\int\prod_{j=1}^{n-1}\uD\Phi_j^+ \uD\Phi_j e^{i\int_{t_0}^t dt L\left(\Phi[\mathbf{X}],\dot{\Phi}[\mathbf{X}]\right)}
\end{split}
\end{equation*}
As a result, we can rewrite the above equation as
\begin{equation}
\label{eq:23}
\fourIdx{*}{}{*}{}{\left<\Phi\middle|TP\{ e^{-i\hat{H}^{(2)}(t-t_0)} \}\middle|\Phi_0\right>}=\int\uuD\Phi^+\uuD\Phi e^{iS^{(cl)}}
\end{equation}
, where 
\begin{equation}
\label{eq:24}
\begin{split}
\uuD\Phi^+\uuD\Phi &\equiv \prod_t \prod_{\mathbf{X}(\sigma)}\uD\Phi^+[\mathbf{X}(\sigma), t]\uD\Phi[\mathbf{X}(\sigma),t]\\
&=\prod_{X(\sigma)}\Phi^+[X(\sigma)]\uD\Phi[X(\sigma)]
\end{split}
\end{equation}
, $L(\Phi[X],\dot{\Phi}[X])$ is the classical lagrangian, and $S^{(cl)}$ is the classical action of string field. As a result, we find that string field can be expressed with a generalized path integral, introducing the path integral measurement as (\ref{eq:24}).

\section{Path integral quantization in light-cone gauge}
We have to use the light-cone string field in order to deal with the relativistic string field. Here we will consider open bosonic string field in light-cone gauge.
Open bosonic string field functional $\Phi_{p^+}[\vec{X},\tau]$ in light-cone gauge satisfies the functional Schrodinger equation\cite{ref5, ref9}
\begin{equation}
\label{eq:25}
i\frac{\partial}{\partial \tau}\Phi_{p^+}[\vec{X},\tau]=\frac{1}{2T}\int d\sigma \left(-\frac{\delta^2}{\delta \vec{X}(\sigma)^2}+T^2\left(\partial_\sigma\vec{X}(\sigma)\right)^2\right)\Phi_{p^+}[\vec{X},\tau]=\hat{h}\Phi_{p^+}[\vec{X},\tau].
\end{equation}
The canonical conjugate momentum of Field functional $\Phi_{p^+}[\vec{X},\tau]$  is determined as 
\begin{equation*}
\Pi_{p^+}[\vec{X},\tau]=\frac{\delta L}{\delta\left(\partial \Phi_{p^+}[\vec{X},\tau]/\partial\tau\right)}=i \Phi^+_{p^+}[\vec{X},\tau]
\end{equation*} 
, and satisfies the following equal-time commutators.
\begin{equation*}
\left[\hat{\Phi}_{p^+}[\vec{X}_1,\tau_1],\hat{\Pi}_{q^+}[\vec{X}_2,\tau_2]\right]_{\tau_1=\tau_2}=i\delta(p^+-q^+)\prod_\sigma \delta\left(\vec{X}_1(\sigma)-\vec{X}_2(\sigma)\right)=i\delta(p^+-q^+)\delta[\vec{X}_1-\vec{X}_2]
\end{equation*} 
, where 
\begin{equation}
\label{eq:26}
L_0=\frac{1}{2}\int_0^\infty dp^+ \int\uD\vec{X}\Phi^+_{p^+}[X]\left(i\frac{\partial}{\partial \tau}-\hat{h}\right)\Phi_{p^+}[\vec{X},\tau]+h.c.
\end{equation}
Field functional operator satisfies the eigenvalue problem
\begin{equation}
\label{eq:27}
\hat{\Phi}_{p^+}[\vec{X},\tau]\left|\Phi\right>=\Phi_{p^+}[\vec{X},\tau]\left|\Phi\right>.
\end{equation}
Defining in this way and passing through the same progresses as above section, the transition amplitude can be expressed with path integral form
\begin{equation}
\label{eq:28}
\fourIdx{*}{}{*}{}{\left<\Phi\middle|TP\{e^{-i\hat{H}^{(2)}(\tau-\tau_0)}\}\middle|\Phi_0\right>}=\int\uuD\Phi^+_{p^+}\uuD\Phi_{p^+}e^{iS_0^{(cl)}}
\end{equation}
, where 
\begin{equation}
\label{eq:29}
\uuD\Phi^+_{p^+}\uuD\Phi_{p^+}\equiv\lim_{n\rightarrow\infty}\uD\Phi^+_{p^+}[\mathbf{X}(\sigma),\tau_1]\uD\Phi_{p^+}[\mathbf{X}(\sigma),\tau_1]\cdots\uD\Phi^+_{p^+}[\mathbf{X}(\sigma),\tau_n]\uD\Phi_{p^+}[\mathbf{X}(\sigma),\tau_n],
\end{equation}
\begin{equation}
\label{eq:30}
S_0^{(cl)}=\int_{\tau_0}^{\tau}L_0\left(\Phi_{p^+},\Phi^+_{p^+}\right)=\int^\tau_{\tau_0}\left[\frac{1}{2}\int^\infty_0\ud p^+\int \uD\vec{X}\Phi^+_{p^+}[X]\left(i\frac{\partial}{\partial \tau}-\hat{h}\right)\Phi_{p^+}[\vec{X},\tau]+h.c\right].
\end{equation}

\section*{Conclusion}
We applied the well-known traditional path integral form to string field from fact that string field is functional field and can be considered as a function with respect to infinite number of variable.

When we are quantizing string field functional, it is not represented as self-conjugate operator. However we use the complete set in Fock space to construct the new complete set of string field state vectors, and on basis of it we introduce the new path integral measurements by space-time slicing approximation and extend Feynman path integral to path integral on functional field. Also we consider how to deal with the non-relativistic and relativistic string fields.

The new path integral measurements by space-time slicing approximation are expressed as (\ref{eq:24}) and (\ref{eq:29}).

Therefore, introducing the generalized path integral measurements, we can find that we get the same quantization as in quantum mechanics and quantum field theory of point particle.

In present paper we consider only the generalized path integral quantization of open bosonic string field, but this idea can be applied to not only real string field but also fermion string field, moreover brane and mass field.

\acknowledgments
I thank all peoples for helping me write this paper as well as  the vice dean of our faculty, Jong Yon-Song.

\end{document}